\parindent = 0 in

\centerline{UNIQUENESS OF ZERO-TEMPERATURE METASTATE IN DISORDERED ISING FERROMAGNETS}
\bigskip
\bigskip
Jan Wehr \footnote{\dag}{wehr@math.arizona.edu} and Aramian Wasielak \footnote{\dag\dag}{Present address:  Fidelity Investments, 200 Seaport Boulevard, Boston, MA 02210}, Department of Mathematics, University of Arizona, Tucson AZ 85721
\bigskip
\bigskip
\centerline{\bf Abstract}
\bigskip
We study ground states of Ising models with random ferromagnetic couplings, proving the triviality of all zero-temperature {\it metastates}.  This unexpected result sheds a new light on the properties of these systems, putting strong restrictions on their possible ground state structure.  Open problems related to existence of interface-supporting ground states are stated and an interpretation of the main result in terms of first-passage and random surface models in a random environment is presented.  
\bigskip
\bigskip
\centerline{\bf Introduction}
\bigskip
Ferromagnetic spin systems with random positive coupling constants are natural disordered versions of the standard Ising model.  Despite the simplicity of their definition, they are difficult to study and basic questions about thembehavior remain unanswered.  This includes the question of ground states---the spin configurations which locally minimize the interaction energy.  The ferromagnetic nature of the system implies that constant spin configurations---all spins equal $+1$ or all equal $-1$---are ground states, but it is not known, in any dimension greater than one, whether other ground states exist.  Nontrivial (nonconstant) ground states would support energetically stable interfaces; one might expect that if they do exist, they  occur in {\it metastates}---translationally covariant probability measures on ground states, defined precisely below.  We show, however, that for a large class of disordered ferromagnets all zero-temperature metastates are supported on the constant spin configurations only.  This is a surprising result which means that other ground states, if they exist, have to be looked for elsewhere.  At the mathematical level, its proof shows that measurability and translational covariance requirements, which are parts of the definition of a metastate, put a very strong restriction on its structure. 
\bigskip
In section 1, after introducing the definitions, we state, prove and discuss the main result, Theorem 1.  Section 2 contains generalizations and extensions, including open problems.
\bigskip
\bigskip
\centerline{\bf Acknowledgements}
\bigskip
The first author was introduced to the question studied here by M. Aizenman.  He would like to thank L.-P. Arguin, C.M. Newman and D. Stein for numerous discussions and for the privilege of their collaboration on a closely related topic, A. Kechris for expertise on measurable selection and A. Fedorenko for renormalization group references.   The research was partially supported by the National Science Foundation grant DMS 0623941.
\bigskip
{\bf 1. Ground states of disordered Ising ferromagnets}
\bigskip
Consider a system of (classical) Ising spins $\sigma_j$ on the lattice ${\bf Z}^d$, interacting by a random Hamiltonian
$$
H_{\bf J}({\bf \sigma}) =- \sum_{|i-j| = 1}J_{ij}\sigma_i \sigma_j. \eqno(1)
$$
Here ${\bf J} = \{J_{ij}: i, j \in{\bf Z}^d, |i - j| = 1\}$ is a realization of an IID family of random variables $J_{ij}$, which are assumed continuously distributed (i.e. without atoms) and positive (ferromagnetic) and ${\bf \sigma}$ is a configuration of ``spin'' values $\sigma_j \in \{-1, 1\}$.
 The expression for $H_{\bf J}$ is formal.  Restricting the summation on its right-hand side to pairs $(i,j)$ intersecting a finite volume $\Lambda$ leads to a well-defined finite-volume energy:
$$
H_{\bf J}({\bf \sigma}) =- \sum_{ij \in \bar{\Lambda}}J_{ij}\sigma_i \sigma_j. \eqno(2)
$$
Here and in the sequel $ij \in \bar{\Lambda}$ means that the bond $ij$ intersects $\Lambda$, i.e. that at least one of the sites $i$ and $j$ belongs to $\Lambda$.  The number of all such bonds (the ``bond volume'' of $\Lambda$) will be denoted by $M(\Lambda)$, while $|\Lambda|$ will be the number of lattice sites in $\Lambda$.  
The difference $H_{\bf J}({\bf \sigma'}) - H_{\bf J}({\bf \sigma})$ is well-defined for any pair of spin configurations such that $\sigma_j = \sigma'_j$ except for finitely many $j$, since only finitely many terms remain in the difference of energies when identical terms are cancelled.  We say in such case that ${\bf \sigma'}$ is a local perturbation of ${\bf \sigma}$. This leads to the fundamental
\bigskip
{\bf Definition:} ${\bf \sigma}$ is a ground state of $H_{\bf J}$ if for all its local perturbations ${\bf \sigma'}$
$$
H_{\bf J}({\bf \sigma'}) - H_{\bf J}({\bf \sigma}) > 0. \eqno(3)
$$
It follows from the continuity of the distribution of the couplings that with probability one the above difference is not equal to zero for any $\sigma$ and its local perturbation $\sigma'$.  
\bigskip
It is clear by ferromagneticity that for every realization of the couplings ${\bf J}$ the uniform spin configurations $\sigma^-_j \equiv -1$ and $\sigma^+_j \equiv 1$ are ground states.  Also, if ${\bf \sigma}$ is a ground state for ${\bf J}$, then so is its global reflection $\{-\sigma_j: j \in {\bf Z}^d\}$.  It follows that ground states come in pairs and that there is always at least one such pair.  
\bigskip
Let us denote by $N$ the number of ground state pairs.  $N$ is a positive integer or infinity (more accurately, an infinite cardinal number, but in this paper different infinite cardinal numbers will not be distinguished).  While {\it a priori} $N$ depends on ${\bf J}$, we have the simple 
\bigskip
{\bf Proposition:} There exists a nonrandom value $N$ (possibly infinite) such that the number of ground state pairs is $N$ with probability one.
\bigskip
Proof:  The group ${\bf Z}^d$ acts naturally on coupling and spin configurations:  for $a \in {\bf Z}^d$
$$
(T_a{\bf J})_{ij} = J_{i-a, j-a} \eqno(4)
$$
and
$$
(T_a{\bf \sigma})_i = \sigma_{i - a}  \eqno(5)
$$
Clearly, ${\bf \sigma}$ is a ground state of $H_{\bf J}$ if and only if $T_a{\bf \sigma}$ is a ground state of $H_{T_a{\bf J}}$, which implies that $N$ is a function of ${\bf J}$, invariant under the translations by lattice vectors defined in (4).  It follows from ergodicity of this group action (or from a version of the Kolmogorov zero-one law) that $N$ is almost surely equal to a constant. ////
\bigskip
Determining $N$ is a major unsolved problem.  It is plausible that $N$ depends only on the dimension $d$ and not on the details of the distribution of $J_{ij}$ (at least for distributions with sufficiently fast tail decay), but this has not been proven.  It was shown in [W] that for any given $d$ and a distribution of the couplings, $N$ equals $1$ or $\infty$.  Several results by C. Newman and his collaborators [N], [HN], [NS1] support the conjecture that $N=1$ for $d=2$, but no proof of it is known.  The analogous conjecture for a half-plane was proven in [WW].  The authors are not aware of any rigorous results in higher dimensions.  Renormalization group calculations [F] suggest that $d=5$ may be the critical dimension above (or perhaps:  at) which nontrivial ground state pairs appear.  
\bigskip
Keeping track of a ground state (or a ground-state pair) of a disordered system while varying the random parameters in the Hamiltonian is difficult.  In essence, one would like to do it, respecting the translational invariance of the model, as expressed by (4) and (5), but it is not clear that such a translationally-covariant ground state can be chosen as a measurable function of ${\bf J}$, and without measurability such an object cannot be used as a tool to study the model.  For some purposes the  concept of a {metastate}, proposed in [AW] and further developed in [N], as well as in [NS2], [NS3], is useful.  It is defined below only for the random ferromagnetic model at zero temperature studied here, but it can be generalized to other disordered systems and to positive temperatures. Note that for a fixed ${\bf J}$ the set of ground states $G_{\bf J}$ of $H_{\bf J}$ is a closed subset of the space of all spin configurations, considered as a countable product of discrete spaces $\{-1,1\}$.  Borel measures on $G_{\bf J}$, invariant under the global reflection ${\bf \sigma} \mapsto -{\bf \sigma}$ are naturally interpreted as measures on the set of ground-state pairs corresponding to this ${\bf J}$.
\bigskip
{\bf Definition:} A zero-temperature metastate is a ${\bf J}$-dependent probability measure $\nu_{\bf J}$ on ground states of $H_{\bf J}$ such that:
\medskip
a) the map ${\bf J} \mapsto \nu_{\bf J}$ is Borel-measurable.
\medskip
b) $\nu_{\bf J}$ varies with ${\bf J}$ in a translationally-covariant way:
$$
\nu_{T_a{\bf J}} = T_a\nu_{\bf J},    \eqno(6)
$$
where the action of $T_a$ on measures is induced by their action on spin configurations (5).
\bigskip
{\bf Remark:} to serve the purposes of [AW], [N] and [ANSW] the metastates defined there were required to satisfy an additional property, describing the behavior of $\nu_{\bf J}$ under local perturbations of the coupling configurations.  No such properties are needed in the present paper.
\bigskip
In some sense, zero-temperature metastates describe behavior of the system at zero-temperature.  One can ask whether all ground states are ``visible'' by looking at metastates only.  More precisely, we have:
\bigskip
{\bf Open question:} Does there exist a metastate, such that for each ${\bf J}$ the support of $\nu_{\bf J}$ is equal to the whole set $G_{\bf J}$?
\bigskip
To the best of the authors' knowledge, the answer is not known.  In particular, the existing general theorems on measurable selection do not seem to apply [Ke].  A positive answer would imply that $N=1$ for all $d$ and for all integrable distributions of $J_{ij}$, as is seen from the following uniqueness of the zero-temperature metastate, the main result of this work:
\bigskip
{\bf Theorem 1:} Assume that the coupling variables $J_{ij}$ have a finite mean.  Every zero-temperature metastate is supported on the uniform ground states.
$$
\nu_{\bf J} = (1 - \alpha)\delta_{{\bf \sigma}^+} + \alpha \delta_{{\bf \sigma}^-}  \eqno(7)
$$
for almost all ${\bf J}$, where $\alpha \in [0,1]$ is constant and $\delta_{\bf \sigma}$ denotes the point mass supported on ${\bf \sigma}$.  
\bigskip
{\bf Proof}:  Let $\Lambda$ be a box, centered around the origin of ${\bf Z}^d$.  For any ${\bf J}$ and any ground state ${\bf \sigma}$ of $H_{\bf J}$, consider the configuration ${\bf \sigma'}$, obtained from ${\bf \sigma}$ by putting all spins in $\Lambda$ to $1$.  This is clearly a local modification of ${\bf \sigma}$, so we have 
$$
H_{\bf J}({\bf \sigma}) - H_{\bf J}({\bf \sigma'}) \leq 0.            \eqno(8)
$$
Retaining only the terms which do not cancel, we can write this as 
$$
H_{{\bf J}, \Lambda}({\bf \sigma}) \leq H_{{\bf J}, \Lambda}({\bf \sigma'}),      \eqno(9)
$$
with the finite-volume energies $H_{{\bf J}, \Lambda}({\bf \sigma})$ defined by restricting the summation to the bonds $ij$, intersecting $\Lambda$, as in (2).  

Denoting, as above, by ${\bf \sigma^+}$ the constant configuration whose all spins equal $+1$ and subtracting $H_{{\bf J}, \Lambda}({\bf \sigma^+})$ from both sides of the above inequality, we get
$$
H_{{\bf J}, \Lambda}({\bf \sigma}) - H_{{\bf J}, \Lambda}({\bf \sigma^+}) \leq H_{{\bf J}, \Lambda}({\bf \sigma'}) - H_{{\bf J}, \Lambda}({\bf \sigma^+}) \leq 2\sum_{ij \in \partial \Lambda} J_{ij}. \eqno(10)
$$
This is the key estimate, in which we used the ferromagnetic nature of the model.  In essence, the above bound estimates the energy contribution from the volume $\Lambda$ to any ground state  by pushing the interfaces between plus and minus spins from the interior of $\Lambda$ to its boundary.  
\medskip
Noting that the left-hand side of (10) equals the sum of $2J_{ij}$ over those bonds $ij$ intersecting $\Lambda$, for which $\sigma_i \sigma_j = -1$ and integrating with respect to $\nu_{\bf J}$, we obtain
$$
{1 \over M(\Lambda)}\sum_{ij \in \bar{\Lambda}}\int \nu_{\bf J}(d{\bf \sigma})I_{\{\sigma_i\sigma_j= -1\}}J_{ij} \leq {1 \over M(\Lambda)}\sum_{ij \in \partial \Lambda}J_{ij}.  \eqno(11)
$$
The left-hand side can be written as
$$
{1 \over M(\Lambda)}\sum_{ij \in \bar{\Lambda}}\phi(ij),                \eqno(12)
$$
with
$$
\phi(ij) = \int \nu_{\bf J}(d{\bf \sigma})I_{\{\sigma_i\sigma_j= -1\}}J_{ij}.                   \eqno(13)
$$
The expression (12) is an ergodic average, which, by the symmetries of the model, converges almost surely to the expected value (i.e. the ${\bf J}$-average of the random variable (13) for any fixed bond $ij$) by a multidimensional version of the Birkhoff ergodic theorem [AK].  The convergence holds in $L^1$ as well, hence, integrating over ${\bf J}$ and passing to the limit, we obtain
$$
E[\phi(ij)] = 0.
$$
Since $\phi(ij) \geq 0$, it follows that $\phi(ij) = 0$ with probability one and this, in turn, implies that for almost every ${\bf J}$ the measure $\nu_{\bf J}$ is supported on configurations ${\bf \sigma}$ with $\sigma_i \sigma_j = 1$.  Intersecting countably many events with probability one, we conclude that for almost every ${\bf J}$ the only configurations in the support of $\nu_{\bf }$ are constant configurations.   This implies (7) with an $\alpha$ which must be equal to a constant for almost all ${\bf J}$ by ergodicity.  The theorem is proven. ////
\bigskip
{\bf Discussion}:  The above proof uses crucially two properties of zero-temperature metastates:  translational covariance and measurable dependence on the coupling realizations.  Theorem 1 shows that these two conditions imply triviality of  metastates.  This by no means rules out existence of nontrivial ground states for typical ${\bf J}$, showing only that the set of such ground states would have to vary with the couplings in a more complicated way then the structure of a metastate allows for.  In the opinion of the authors this calls for further study of the ground states of disordered ferromagnets, as a foundation for understanding their statistical mechanics. 
\bigskip
\bigskip
{\bf 2.  Corollaries and extensions}
\bigskip
\bigskip
A relation between ground states of disordered ferromagnets and minimal hypersurfaces in random environment is well-known.  Its brief discussion sufficient for our purposes can be found in [W].  In essence, one considers hypersurfaces built from $d-1-$dimensional cells, dual to bonds of ${\bf Z}^d$.  A hypersurface is assigned a formal energy, equal to the sum of the $J_{ij}$ dual to its cells.  Similarly to spin configurations, if two hypersurfaces are local modifications of each other, the difference of their energies is well-defined.  A hypersurface is called minimal if the energy difference between it and any of its local modifications is negative.  As for the spin model, one can define a metastate as a map, which to every ${\bf J}$ assigns a probability measure on minimal hypersurfaces in a measurable and translationally-covariant way.  The following theorem has a proof very similar to the proof of Theorem 1:
\bigskip
{\bf Theorem 2:}  Under the same assumptions as in Theorem 1, there is no metastate supported on minimal hypersurfaces. ////
\bigskip
In two dimensions, hypersurfaces are lines and theorem 2 says that there is no metastate supported on minimal lines.  It should be stressed that this does not imply that minimal lines do not exist---it is an open question.  Minimal lines (as opposed to $d-1$-dimensional hypersurfaces) can be also considered when $d>2$.  Again, a similar argument proves the following.
\bigskip
{\bf Theorem 3:} Under the same assumptions as in Theorem 1, there is no metastate supported on minimal lines. ////
\bigskip
Clearly, for even greater generality, minimal hypersurfaces of arbitrary dimensions can be studied.  We will not state the obvious generalization.
\bigskip
We return to disordered Ising ferromagnets and generalize 
Theorem 1 to a class of nonintegrable distributions.
\bigskip
{\bf Theorem 4}:
Suppose that $J_{ij}$ are IID random variables satisfying the
condition
$$
{1 \over M_L} \sum_{ij \in
\partial \Lambda_L} J_{ij} \buildrel \rm d \over \to 0.
$$
Here and in the sequel $\buildrel \rm d \over \to$ denotes
convergence in distribution (which in the case of convergence to a constant limit is
equivalent to convergence in probability).
Then the conclusion of Theorem 1 holds.
\bigskip
{\bf Proof}:
We use the same notation as in the proof of Theorem 1.  The idea is to replace the (now non-integrable) random variables $J_{ij}$ by bounded ones.   Let $f$ be a continuous function which is
strictly increasing, concave and bounded on $[0,
\infty)$ such that $f(0) = 0$ (e.g. $f(x) = \arctan{x}$).  In the following string of equalities and inequalities we use (in this order):  the ergodic theorem, concavity of $f$, its monotonicity, together with inequality (11), the definition of convergence in distribution and the bounded convergence theorem. 
$$\eqalign{
{\bf E}\big[\int\nu_{\bf J}(d\sigma)f(J_{ij}I_{\{\sigma_i\sigma_j = -1\}})\big] &= \cr
\lim_{L \to \infty}{1 \over M_L}\sum_{ij \in \Lambda_L}\int\nu_{\bf J}(d{\bf \sigma})f(J_{ij}I_{\{\sigma_i\sigma_j = -1\}}) &=  \cr
\lim_{L \to \infty}\int\nu_{\bf J}(d\sigma){1 \over M_L}\sum_{ij \in \Lambda_L}f(J_{ij}I_{\{\sigma_i\sigma_j = -1\}}) &\leq \cr
\lim_{L \to \infty}\int\nu_{\bf J}(d\sigma)f\big({1 \over M_L}\sum_{ij \in \Lambda_L}J_{ij}I_{\{\sigma_i\sigma_j = -1\}}\big) &\leq \cr
\lim_{L \to \infty}\int\nu_{\bf J}(d\sigma)f\big({1 \over M_L}\sum_{ij \in \partial \Lambda_L}J_{ij}\big) =  0 \cr
}
$$
The theorem follows ////
\bigskip
{\bf Remark:}  The condition
$$
{1 \over\Lambda_L} \sum_{ij \in
\partial \Lambda_L} J_{ij} 
\buildrel \rm d \over \to 0,
$$
assumed in the Theorem 2, is satisfied by any distribution in the
domain of attraction of the stable law whose characteristic
function $f$ satisfies
$$
 \log f(u) = i u + m_{1}
\int_{0}^{\infty} \left( e^{iux} - 1 \right)
{dx \over x^{1+\alpha}} + m_{2} \int_{-\infty}^{0} \left( e^{iux}
- 1 \right) {dx \over|x|^{1+\alpha}}.
$$
for ${d-1 \over d} < \alpha < 1$, $m_{1} \geq 0$, and $m_{2} \geq
0$. More precisely, the above formula gives the characteristic
function of a stable distribution of index $\alpha < 1$,
[B], page 204, and in [Z], page 80, it
is shown that all (appropriately normalized) stable distributions
supported on ${ \bf R_+}$ are described this way. These
distributions are nonintegrable, i.e. a random variable with such
distribution has infinite first moment). If $X, X_1, \dots, X_n$
are independent random variables with such distribution, we have
$$
{X_1 + \dots + X_n \over n^{1 \over \alpha}} \buildrel \rm d \over
= X
$$
and, consequently, if $J_{ij}$ has such distribution, we obtain
$$
{1 \over L^{d}} \sum_{b \in \partial \Lambda_L} J_{ij} \buildrel
\rm d \over \to 0
$$
whenever $\alpha > {d-1 \over d}$.  The same is true for any
distribution $\mu$ in the domain of attraction of the stable law, in the sense that
$$
{X_1 + \dots + X_n \over n^{1 \over \alpha}} \buildrel \rm d \over
\to X,
$$
where the distribution of $X$ is the stable law defined above and $X_1, \dots, X_n, \dots$ are IID
random variables with the distribution $\mu$. We have thus
identified a class of nonintegrable passage time
distributions to which Theorem 2 applies---those in the domain of
attraction of stable laws whose tail probabilities decay
sufficiently fast.  With appropriate modifications, the above discussion
applies to models with weakly dependent passage times.  We will
not discuss details here.
\bigskip
{\bf Remark:}   A one-sided line is a line on the lattice dual to ${\bf Z}^d$, which starts at a certain point.  Its minimality is defined in the above way, as stability under local perturbations that preserve the starting point. Unlike the minimal lines discussed here, the {\it one-sided} minimal lines are known to exist; theire existence follows from a simple diagonal choice (or:  compactness) argument.  Detailed properties of such lines in this and related models have been under intense study in recent years.  See [Ch, ADH, LaGW].
\bigskip
{\bf Remark:}  In [W] it was proven that the number of ground state pairs, $N$ is $1$ or infinity.  This follows from Theorem 1:  $1 < N < \infty$ would imply that the normallized counting measure on the $N$ ground state pairs is a metastate not supported on the constant configurations only.
\bigskip
{\bf Remark:} The proof and even the statement of Theorem 1 rely crucially on the  ferromagneticity assumption.  No analogous theorems are available for {\it spin glass} models, where this assumption does not hold.  See [AD, ADNS, ANSW] for some recent results on such models, at zero, as well as at positive temperatures.
\bigskip
{\bf Remark:} At positive temperatures, metastates are defined analogously, as ${\bf J}$-dependent probability measures on Gibbs states of the interaction $H_{\bf J}$.  Two Gibbs states $\mu^-_{\bf J}$ and $\mu^+_{\bf J}$ can be constructed using uniform $-1$ and $+1$ boundary conditions. They provide a positive-temperature analog of the uniform ground states ${\bf \sigma^-}, {\bf \sigma^+}$.  Note however that, unlike ${\bf \sigma^{\pm}}$, the states ${\bf \mu}^{\pm}_{\bf J}$ do depend on ${\bf J}$.  It would be interesting to prove an analog of Theorem 1 for positive temperatures, proving that every metastate is supported on these special Gibbs states.  This does not appear to be straightforward.
\bigskip
\bigskip
\centerline{\bf References}
\bigskip
[AW] Aizenman, M., Wehr, J.:  Rounding effects of quenched randomness on first-order phase transitions.
Comm. Math. Phys. 130, 489–528 (1990)
\medskip
[AK] Akcoglu, M., Krengel, J.:  Ergodic Theorems for Superadditive Processes. J. Reine Angew. Math. 323, 53-67 (1981)
\medskip
[AD] Arguin, L.-P., Damron, M.:  On the Number of Ground States in the Edwards-Anderson Spin Glass Model. Ann. Inst. H. Poincar\'e Probab. Statist. 50, number 1 (2014)
\medskip
[ADMS] Arguin, L.-P., Damron, M., Newman, C., Stein D.:  Uniqueness of Ground States for Short-Range Spin Glasses in the Half-Plane.  Comm. Math. Phys. 300 (2010) 
\medskip
{ADH] Auffinger, A., Damron, M., Hanson, J.:  Limiting geodesic for first-passage percolation on subsets of ${\bf Z}^2$. to appear in Ann. Appl. Prob. (2014)
\medskip
[ANSW] Arguin, L.-P., Newman, C., Stein, D., Wehr, J.:  Fluctuation Bounds For Interface Free Energies in Spin Glasses.  To appear in J. Stat. Phys. (2014)
\medskip
[B] Breiman, L.:  Probability.  Addison-Wesley (1968)
\medskip
[Ch] Chatterjee, S:  The universal relation between scaling exponents in first-
passage percolation. Ann. of Math. (2), 177, 663–697 (2013)
\medskip
[F] Fedorenko, A., private communication
\medskip
[HN] Howard, C.D., Newman, C.M.:  Geodesics And Spanning Tees For Euclidean First­Passage Percolaton.  Ann. Probab. 29, no. 2, 577-623 (2001)
\medskip
[K] Kechris, A.:  private communication
\medskip
[LaGW} LaGatta, T., Wehr, J.:  Geodesics of Random Riemannian Metrics.  Commun. Math. Phys. 327(1), 187-241 (2014)
\medskip
[N] Newman, C.M.:  Topics in disordered systems.  Springer (1997)
\medskip
[NS1] Newman, C.M., Stein, D.L.:  Ground state structure in a highly disordered spin glass model.  J. Stat. Phys. 82, 1113-1132 (1996)
\medskip
[NS2] Newman, C.M., Stein, D.L.:  Interfaces and the question of regional congruence in spin glasses.
Phys. Rev. Lett. 87, 077201 (2001)
\medskip
[NS3] Newman, C.M., Stein, D.L.: Are There Incongruent Ground States in 2D EdwardsAnderson Spin
Glasses? Comm. Math. Phys. 224, 205–218 (2001)
\medskip
[W] Wehr, J.:  On the number of infinite geodesics and ground states in disordered systems.  J. Stat. Phys. 87, 1-2, pp 439-447 (1997)
\medskip
[WW] Wehr, J, Woo, J.:  Absence of geodesics in first-passage percolation on a half-plane.  Ann. Probab. 26, no. 1, 358-367 (1998)
\medskip
[Z] Zolotarev, V.M.:  One-dimensional stable distributions.  AMS (1986)

\bigskip

\bye